\documentclass[showpacs,aps,graphicx,twocolumn]{revtex4}
\usepackage{amssymb}
\usepackage{amsmath}
\usepackage{graphicx}
\usepackage{dcolumn}
\usepackage{bm}
\usepackage{array}

\begin{document}

\title{Unidirectional transport of wave packets through tilted discrete
breathers in nonlinear lattices with asymmetric defects\footnote{Published in Phys. Rev. E \textbf{94}, 032216 (2016).}}

\author{Xiao-Dong Bai,$^1$ Boris A. Malomed,$^{2,3}$ and Fu-Guo Deng$^{1,}$\footnote{Corresponding author: fgdeng@bnu.edu.cn}}

\address{
$^1$Department of Physics and Applied Optics Beijing Area Major
Laboratory, Beijing Normal University, Beijing 100875, China\\
$^2$Department of Physical Electronics, School of Electrical
Engineering, Faculty of Engineering, Tel Aviv University, Tel Aviv
69978, Israel\\
$^3$Laboratory of Nonlinear-Optical Informatics, ITMO University,
St. Petersburg 197101, Russia}

\begin{abstract}
We consider the transfer of lattice wave packets through a tilted discrete
breather (TDB) in opposite directions in the discrete nonlinear Schr\"{o}%
dinger model with asymmetric defects, which may be realized as a
Bose-Einstein condensate trapped in a deep optical lattice, or as
optical beams in a waveguide array. A unidirectional transport
mode is found, in which the incident wave packets, whose energy
belongs to a certain interval between full reflection and full
passage regions, pass the TDB only in one direction, while, in the
absence of the TDB, the same lattice admits bi-directional
propagation. The operation of this mode is accurately explained by
an analytical consideration of the respective energy barriers. The
results suggest that the TDB may emulate the unidirectional
propagation of atomic and optical beams in various settings. In
the case of the passage of the incident wave packet, the
scattering TDB typically shifts by one lattice unit in the
direction from which the wave packet arrives, which is an example
of the \textit{tractor-beam} effect, provided by the same system,
in addition to the rectification of incident waves.
\end{abstract}

\author{}
\maketitle



\section{Introduction}

Time-periodic spatially localized excitations, called discrete breathers
(DBs) \cite{AJSievers1998,SFlach1998,Panos}, are supported by diverse
nonlinear lattice media, such as arrays of micromechanical cantilevers \cite%
{MSato2006}, coupled Josephson junctions \cite{ETrias2000,AVUstinov2003} and
optical waveguides \cite{DNChr,RMorandotti1999,HSEisenberg1998,Kobelke} (in
the latter case, the evolution variable is not time but the propagation
distance), antiferromagnets \cite{UTSchwarz1999,MSato2004}, Bose-Einstein
condensates (BECs) \cite{ATrombettoni2001,BEiermann2004,Konotop} and
Tonks-Girardeau \cite{Kolomeisky1992} or superfluid fermionic \cite%
{JuKuiXue2008} gases fragmented and trapped in deep optical lattices (OLs),
as well as some dissipative systems \cite{MGClerc2011,PCMatthews2011}. DBs
provide means for energy concentration and transport in waveguides \cite%
{SFlach1998}, polaronic materials \cite{JDAndersen1993}, long biological
molecules \cite{MPeyrard1993}, and other settings \cite{Panos}.

DBs may represent attractors in dissipative systems \cite%
{SFlach1998,PJMartinez2003,RSMacKay1998} and self-trapped localized
stationary modes in conservative nonlinear lattices \cite%
{GPTsironis1996,RLivi2006,GSNg2009} (we keep acronym DB\ in those stationary
states too, although a more appropriate name for them may be merely discrete
solitons, as they do not feature breathing dynamics). Collisions of moving
solitons or phonons (linear wave packets) with stationary DBs were studied
too \cite{RLivi2006,GSNg2009,RFranzosi2007,XDBai2012}. It was found that, if
the amplitude of the incident soliton is too small, it bounces back from the
DB. Beyond a specific threshold amplitude, a part of the norm of the
incident solitary pulse may be transmitted through the DB. In particular,
this transport property is vital for matter-wave interferometry \cite{TSchumm2005} and quantum-information processing \cite{DJaksch1999,GKBrennen1999,JKPachos2003,A.Kay2006}.

Previous works were mainly dealing with symmetric DBs in perfect
lattices, the corresponding transfer mechanism being symmetric
with respect to the direction of motion of the incident
excitation. On the other hand, unidirectional propagation of waves
in specially devised systems has drawn much attention too (see
Ref. \cite{Lepri}, and references therein). In particular, it was
experimentally demonstrated that a periodically poled waveguide
may serve as an optical diode \cite{Assanto}. Further, it was
predicted that a combination of a Bragg grating with a periodic
lattice built of gain and loss elements \cite{Bragg-uni}, as well
as a periodically
structured metamaterial \cite{meta}, a chain of driven ultracold atoms \cite%
{atomic-chain-uni}, and chains of coupled microcavities \cite{Fabio} may
give rise to nonreciprocal light transmission. Starting from early work \cite%
{Scalora}, it was also demonstrated theoretically that photonic crystals
(PCs) with edge modes \cite{Ablowitz}, PCs with built-in periodic lattices
of defects \cite{PhotCryst-uni}, second-harmonic-generating PCs \cite{VKVK},
and quasiperiodic PCs \cite{Fabio2}, may produce a similar effect. Other
theoretically predicted and experimentally realized possibilities for the
unidirectional light transmission are offered by $\mathcal{PT}$-symmetric
photonic lattices \cite{PT-uni}. Asymmetric propagation of microwaves \cite%
{micro-uni} and electric signals \cite{electric-uni} was observed,
respectively, in appropriately built electromagnetic crystals and electric
transmission lines.

Realistic lattices always contain imperfections (defects). In many cases,
they dramatically affect the properties of lattices carrying onsite
nonlinearities, resulting in a number of new phenomena, such as Anderson
localization \cite{PWAnderson1958,JBilly2008}, intraband discrete breathers
\cite{GKopidakis2000}, and the disorder-induced Bose-glass phase \cite%
{LFallani2007}. The influence of disorder on the lattice transport
was widely investigated too. It was demonstrated, in various
settings, that the interplay of the disorder with self-defocusing
nonlinearity may replace the localization by subdiffusion
\cite{AIomin2010}. More subtle dynamical regimes, represented by
infinite-dimensional Kolmogorov-Arnold-Moser tori, are possible
too \cite{MJohansson2010}.

Unlike those works, our subject here is modification of the
lattice transport by DBs pinned to local defects. In particular,
if defects have an asymmetric shape, they asymmetrically deform
pinned lattice solitons (alias discrete breathers)
\cite{GKopidakis2000}, making them \textit{tilted} discrete
breathers (TDBs). In this work, we consider a one-dimensional
discrete nonlinear Schr\"{o}dinger (DNLS) model with asymmetric
defects and demonstrate that, while the underlying lattice admits
bidirectional transmission of linear waves, the transmission may
be made nonreciprocal for waves hitting a pinned TDB from opposite
directions. As a result, we find a ``diodelike" transport mode,
where the incident wave packets can pass through the TDB only in
one direction, provided that the packet's energy falls into an
appropriate interval between cases of full reflection and full
passage (for small and large energies, respectively). Thus, while
a localized asymmetric defect cannot directly induce the
unidirectional transmission, it may create such a regime with the
help of the pinned TDB. This is possible because the scattering of
waves on a soliton in a nonlinear system is essentially different
from the scattering on a defect in the linear counterpart of the
same system. In particular, the nonlinearity gives rise to the
interaction of the wave with its own complex-conjugate
counterpart, which does not happen in the linear limit. The diode
mode can be controlled by varying parameters of the underlying
defect, as well as the amplitude (i.e., energy) of the pinned TDB.

The paper is organized as follows. In Sec. \ref{sec2}, we introduce the
model and produce tilted discrete breathers for dilute gases of bosonic
atoms trapped in the deep OL. Unidirectional transport of wave packets
through the TDBs is reported in Sec. \ref{sec3}. Conclusions are presented
in Sec. \ref{sec4}. Additional technical details and limit cases are
presented in the Appendix.

\section{The model and tilted discrete breathers}

\label{sec2} We start by taking the Bose-Hubbard Hamiltonian, which is the
basic model that captures the dynamics of a dilute gas of bosonic atoms in
the deep OL. In the mean-field approximation, the Hamiltonian for the
disordered OL is \cite{ZYSun2015,DJaksch1998}
\begin{equation}
\mathcal{H}\!=\!\sum_{n=1}^{M}\left( \frac{U}{2}|\psi
_{n}|^{4}\!+\!\varepsilon _{n}|\psi _{n}|^{2}\right) -\!\frac{J}{2}%
\!\sum_{n=1}^{M-1}\!\left( \psi _{n}^{\ast }\psi _{n+1}\!+\mathrm{c.c.}%
\right) ,  \label{eq1}
\end{equation}%
where $n$ $=1,\ldots,M$ is the discrete coordinate of lattice
sites, $\psi _{n}$ is the on-site wave function, $J$ is the
amplitude of the inter-site hopping, and c.c. stands for the
complex conjugate. Further, $U=4\pi \hbar ^{2}a_{s}V_{d}/m$ is the
strength of the collisional on-site interactions, where $V_{d}$ is
the effective mode volume of each site, $m$ is the atomic mass,
and $a_{s}$ is the \textit{s}-wave atomic scattering length. We
consider repulsive interactions, with $a_{s}>0$. Lastly,
$\varepsilon _{n}$ is on-site energy which accounts for the
disorder.

The scaled DNLS equation following from Hamiltonian (\ref{eq1}) is
\begin{equation}
i\frac{d\psi _{n}}{dt}\;=\;\lambda \left\vert \psi _{n}\right\vert
{}^{2}\psi _{n}\!+\varepsilon _{n}\psi _{n}\!-\!\frac{1}{2}\left( \psi
_{n-1}\!+\psi _{n+1}\right) ,  \label{eq2}
\end{equation}%
with the normalization defined by $\lambda \equiv U/J$ and $t\equiv JT$,
where $T$ is time measured in physical units. Additional rescaling makes it
possible to fix the nonlinearity coefficient, which is set below to be $%
\lambda =3$, unless stated otherwise. Equation (\ref{eq2}) conserves
Hamiltonian (\ref{eq1}) and the total norm, $\sum_{n=1}^{M}|\psi _{n}|^{2}$.

In addition to BEC, Eq. (\ref{eq2}) applies to the light propagation in
arrays of nonlinear optical waveguides \cite{DNChr,HSEisenberg1998}, with $t$
replaced by the propagation distance ($z$), $\lambda $ being the effective
Kerr coefficient, and the inter-site coupling coefficient normalized to be $%
1 $. In this case, $-\varepsilon _{n}$ is proportional to deviation of the
local refractive index from its average value, and $\psi _{n}(z)$ is the
on-site amplitude of the electromagnetic wave in the spatial domain.

The transport of excitations through the DB is mainly determined by the
dynamics at three sites on which a narrow DB is located. Therefore, we first
address the trimer model ($M=3$) with system (\ref{eq2}) truncated to
\begin{equation}
\begin{split}
i\frac{d\psi _{1}}{dt}\;& =\;\lambda \left\vert \psi _{1}\right\vert
{}^{2}\psi _{1}+\varepsilon _{1}\psi _{1}-\frac{1}{2}\psi _{2}, \\
i\frac{d\psi _{2}}{dt}\;& =\;\lambda \left\vert \psi _{2}\right\vert
{}^{2}\psi _{2}+\varepsilon _{2}\psi _{2}-\frac{1}{2}\left( \psi _{1}+\psi
_{3}\right) , \\
i\frac{d\psi _{3}}{dt}\;& =\;\lambda \left\vert \psi _{3}\right\vert
{}^{2}\psi _{3}+\varepsilon _{3}\psi _{3}-\frac{1}{2}\psi _{2},
\end{split}
\label{eq3}
\end{equation}%
subject to normalization $\sum_{n=1}^{3}\left\vert \psi _{n}\right\vert
{}^{2}=1.$ First, DB\ solutions are looked for as
\begin{equation}
\psi _{n}(t)\;=\;A_{n}(t)\exp \left( -\text{$i\mu t$}\right) ,
\label{solution}
\end{equation}%
with real amplitudes $A_{n}$ and chemical potential $\mu $ (in optics, $-\mu
$ is the propagation constant). Localized (self-trapped) discrete modes in
the system with the repulsive on-site nonlinearity can be found in the
\textit{staggered} form, i.e., with alternating signs of the amplitudes at
adjacent sites \cite{Panos},
\begin{equation}
A_{1}>0,\;\;\;A_{2}=-\sqrt{1-A_{1}^{2}-A_{3}^{2}}<0,\;\;\;A_{3}>0
\label{123}
\end{equation}%
(written so as to comply with the normalization condition). Then, the
substitution of Eq. (\ref{solution}) into Eq. (\ref{eq3}) for $\psi _{2}$
yields%
\begin{equation}
\mu =\varepsilon _{2}+\lambda \left( 1-A_{1}^{2}-A_{3}^{2}\right) +\left(
A_{1}+A_{3}\right) /\sqrt{1-A_{1}^{2}-A_{3}^{2}}.  \label{mu}
\end{equation}%
As $\varepsilon _{2}$ can be absorbed into a shift of $\mu $ in Eq. (\ref{mu}%
), we set $\varepsilon _{2}=0$ below. Two remaining equations for $\psi
_{1,3}$ in Eq. (\ref{eq3}) are%
\begin{equation}
2A_{1,3}\!\!\left[ \varepsilon _{1,3}\!\!+\lambda \!\!\left(
2A_{1,3}^{2}\!\!+\!\!A_{3,1}^{2}-1\right) \!\right] \!=\frac{%
2A_{1,3}^{2}\!\!+\!\!A_{3,1}^{2}\!\!+\!\!A_{1}A_{3}\!\!-\!\!1}{\sqrt{%
1\!\!-\!\!A_{1}^{2}\!\!-\!\!A_{3}^{2}}},  \label{eq4}
\end{equation}%
where $\mu $ is eliminated with the help of Eq. (\ref{mu}).

\begin{figure}[tbp]
\includegraphics[width=8cm]{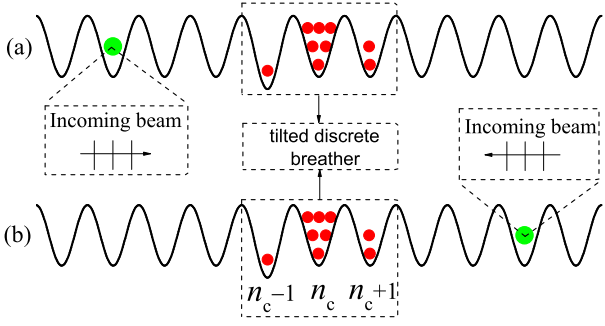}
\caption{The transport scheme under the consideration. Large green
dots denote identical incident wave packets coming from the left
in (a) and the right
in (b). The middle part schematically shows the TDB, where $\protect%
\varepsilon _{n}=0$, except for $\protect\varepsilon _{n_{c}-1}<0$. }
\label{Fig1}
\end{figure}

From Eq. (\ref{eq4}), one can obtain different types of stationary
solutions. For the symmetric configuration with $\varepsilon
_{1}=\varepsilon _{3}$, the stationary states correspond to ordinary
symmetric DBs, which have been investigated in detail, including their
formation \cite{RLivi2006,GSNg2009} and transport properties \cite%
{RFranzosi2007,XDBai2012,HHennig2010}. On the other hand, the setting with $%
\varepsilon _{1}\neq \varepsilon _{3}$ gives rise to TDB, as schematically
shown in the middle part of Fig. \ref{Fig1}, where $n_{c}-1$, $n_{c}$, and $%
n_{c}+1$ may be regarded as $1$, $2$, and $3$, respectively, in the trimer
model.

\begin{figure}[tbp]
\includegraphics[width=8.0cm]{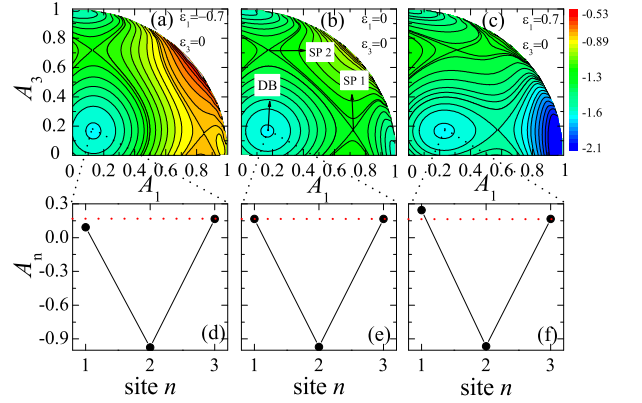}
\caption{(a)-(c) Contour plots of energy given by Eq.
(\protect\ref{eq7}) for different $\protect\varepsilon _{1}$ with
fixed $\protect\lambda =3$ and $\protect\varepsilon _{3}=0$.
(d)-(f) The structure of the DB corresponding to panels (a)-(c),
respectively. SP1 and SP2 denote the saddle points.} \label{Fig2}
\end{figure}

Here, our aim is to investigate the transport of lattice waves through the
TDB, for the wave packets arriving from the opposite directions (see Fig. %
\ref{Fig1}). To this end, we consider the system's Hamiltonian for
configurations taken in the form of Eqs. (\ref{solution}) and
(\ref{123}), which must be taken with the opposite overall sign,
due to the above-mentioned staggering transformation:
\begin{equation}
H=-\frac{\lambda }{2}\!\left( A_{1}^{4}\!+\!A_{2}^{4}\!+\!A_{3}^{4}\right)
\!-\!\left( \varepsilon _{1}A_{1}^{2}\!+\!\varepsilon _{3}A_{3}^{2}\right)
\!-\!\left( A_{1}\!+\!A_{3}\right) \!A_{2}.  \label{eq7}
\end{equation}%
The effect of $\varepsilon _{n}$ on the Hamiltonian, considered as
a function of $A_{1}$ and $A_{3}$, is shown in Figs.
\ref{Fig2}(a)-\ref{Fig2}(c), which exhibits three minima separated
by two saddle points SP1 and SP2. The minima refer to the DBs,
whose structure is displayed in Figs. \ref{Fig2}(d)-\ref{Fig2}(f)
(the trimer model per se was studied in detail in Ref. \cite{GKopidakis2000}). Analytical solutions for SP1, SP2, and DB can be obtained from Eq. (\ref%
{eq4}) in the limit of large $\lambda $ and $\varepsilon _{1}, \varepsilon
_{3}\rightarrow 0$. Similar to the results of Ref. \cite{GKopidakis2000},
the respective energies of SP1 and SP2, $E_{\mathrm{thr}1,2}$, which play
the role of thresholds for the perturbed dynamics of the trimer (see below),
are
\begin{eqnarray}
E_{\mathrm{thr}1}\! &=&\!-\!\frac{1}{2}-\frac{\lambda }{4}-\frac{1}{4\lambda
}+\frac{1}{4\lambda ^{2}}-\frac{1}{4\lambda ^{3}}+\frac{9}{16\lambda ^{4}}
\notag \\
&&-\varepsilon _{1}\!\!\left( \!\frac{1}{2}\!+\!\!\frac{1}{2\lambda ^{3}}%
\!-\!\frac{5}{8\lambda ^{4}}\!\right) \!\!+\!\varepsilon _{1}^{2}\!\!\left(
\!\!\frac{1}{4\lambda }\!+\!\!\frac{1}{4\lambda ^{2}}\!+\!\!\frac{1}{%
16\lambda ^{3}}\!+\!\!\frac{3}{16\lambda ^{4}}\!\!\right)  \notag \\
&&-\varepsilon _{1}^{3}\left( \frac{1}{8\lambda ^{3}}+\frac{1}{16\lambda ^{4}%
}\right) \!\!-\!\varepsilon _{3}\!\!\left( \!\!\frac{1}{2\lambda ^{2}}\!-\!%
\frac{1}{\lambda ^{3}}\!+\!\frac{1}{2\lambda ^{4}}\!\!\right) \!,
\label{eq9}
\end{eqnarray}%
\begin{eqnarray}
E_{\mathrm{thr}2}\! &=&\!-\!\frac{1}{2}-\frac{\lambda }{4}-\frac{1}{4\lambda
}+\frac{1}{4\lambda ^{2}}-\frac{1}{4\lambda ^{3}}+\frac{9}{16\lambda ^{4}}
\notag \\
&&-\varepsilon _{3}\!\!\left( \!\frac{1}{2}\!+\!\!\frac{1}{2\lambda ^{3}}%
\!-\!\frac{5}{8\lambda ^{4}}\!\right) \!\!+\!\varepsilon _{3}^{2}\!\!\left(
\!\!\frac{1}{4\lambda }\!+\!\!\frac{1}{4\lambda ^{2}}\!+\!\!\frac{1}{%
16\lambda ^{3}}\!+\!\!\frac{3}{16\lambda ^{4}}\!\!\right)  \notag \\
&&-\varepsilon _{3}^{3}\left( \frac{1}{8\lambda ^{3}}+\frac{1}{16\lambda ^{4}%
}\right) \!\!-\!\varepsilon _{1}\!\!\left( \!\!\frac{1}{2\lambda ^{2}}\!-\!%
\frac{1}{\lambda ^{3}}\!+\!\frac{1}{2\lambda ^{4}}\!\!\right) \!.
\label{eq10}
\end{eqnarray}

In the case of $\varepsilon _{1}=\varepsilon _{3}=0$, which corresponds to
Fig. \ref{Fig2}(b), the discrete solitons corresponding to SP1, SP2, and the
DB are symmetric, as shown in Fig. \ref{Fig2}(e). On the other hand, the
asymmetric setting, with
\begin{equation}
\varepsilon _{1}<0,\;\;\;\;\;~\varepsilon _{2}=\varepsilon _{3}=0,
\label{asymm}
\end{equation}
with $\varepsilon _{1}=-0.7$, is represented by Fig. \ref{Fig2}(a). In this
case, the discrete soliton becomes a TDB, as shown in Fig. \ref{Fig2}(d),
where the norm (alias power, in terms of optics) localized at site $1$ is
smaller than in the symmetric case. SP2 remains nearly the same as in the
symmetric case, while SP1 moves away from the TDB. The relationship between
energy thresholds in this case is $E_{\mathrm{thr}1}>E_{\mathrm{thr}2}$.
Similarly, we set $\varepsilon _{3}=0$ and $\varepsilon _{1}=0.7>0$ in Fig. %
\ref{Fig2}(c), which again gives rise to a TDB, with the norm
(power) at site $1$ larger than that in the symmetric case, see
Fig. \ref{Fig2}(f). In this case, SP2 remains nearly unaffected,
while SP1 moves towards the TDB,
and the relationship between the energy thresholds is $E_{\mathrm{thr}1}<E_{%
\mathrm{thr}2}$. Note that, in the limit of $\varepsilon
_{1}\rightarrow -\infty $, SP1 will disappear and value
$E_{\mathrm{thr}1}$ cannot be reached, as shown in Fig.
\ref{SFig1}(a) in the Appendix.

\begin{figure}[tbp]
{\LARGE \includegraphics[width=8.0cm]{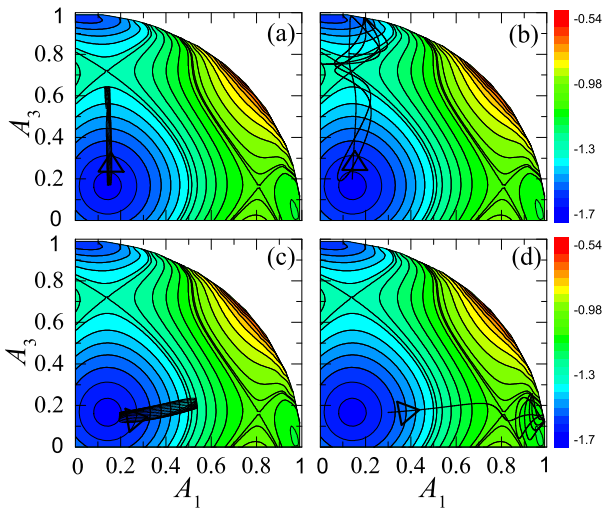} }
\caption{Dynamics of the trimer initiated by the perturbations applied at
site $3$ [(a),(b)] or at site $1$ [(c),(d)]. Here $E_{\mathrm{thr}1}=-1.01787>E_{%
\mathrm{thr}2}=-1.29859$ correspond to SP1 and SP2, respectively. (a) $%
H=-1.319<E_{\mathrm{thr}2}$, and the orbit cannot climb over SP2. (b),(c) $E_{%
\mathrm{thr}2}<H=-1.263<E_{\mathrm{thr}1}$, hence the orbit can pass SP2,
but cannot pass SP1. (d) $H=-1.011>E_{\mathrm{thr}1}$, and the orbit can
climb over the SP1. Perturbation parameters in Eq. (\protect\ref{pert}) are $%
\protect\delta A_{3}=0.004528$ in (a) and $\protect\delta A_{3}=0.032866$ in
(b). Similarly, perturbations applied at site $1$ are $\protect\delta %
A_{1}=0.055125$ in (c) and $\protect\delta A_{1}=0.158584$ in (d). In all
the cases, $\protect\lambda =3$, $\protect\varepsilon _{1}=-0.5$, $\protect%
\varepsilon _{3}=0$, $A_{1}^{\mathrm{TDB}}=0.1412,A_{3}^{\mathrm{TDB}%
}=0.1664 $, and $\protect\delta \protect\theta =\protect\pi $.}
\label{Fig3}
\end{figure}

\section{Unidirectional transport of wave packets}

\label{sec3} To study the transport of matter waves through the TDB in
opposite directions, we first consider the incident wave coming from the
right. In terms of the natural representation, $\psi _{n}(t)=A_{n}e^{i\theta
_{n}}$, where $A_{n}$ and $\theta _{n}$ are amplitudes and phases at site $n$%
, the corresponding initial conditions for the trimer approximation may be
considered as the stationary TDB disturbed at site $3$:
\begin{equation}
\psi {(t=0)}=\left\{ A_{1}^{\mathrm{TDB}},A_{2},(A_{3}^{\mathrm{TDB}}+\delta
A_{3})e^{i\delta \theta }\right\} ,  \label{pert}
\end{equation}%
with $A_{2}=-\sqrt{1-|\psi _{1}|^{2}-|\psi _{3}|^{2}}$, perturbation
amplitude $\delta A_{3}$ added at site $3$, and the corresponding phase $%
\delta \theta $. The resulting evolution of the trimer system, initiated by
the perturbation, is shown by black trajectories in Fig. \ref{Fig3} for
fixed $\lambda =3$, $\varepsilon _{1}=-0.5$, and $\varepsilon _{3}=0$, so
that $E_{\mathrm{thr}1}=-1.01787>E_{\mathrm{thr}2}=-1.29859$. If the
perturbation is small, the total energy of the system, $H$, remains smaller
than $E_{\mathrm{thr}2}$, hence the orbit is not able to climb over saddle
point SP2, while the TDB remains stable. That is, for this case nothing is
transferred to site $1$, as shown in Fig. \ref{Fig3}(a). If the perturbation
is large, so that $H>E_{\mathrm{thr}2}$, the orbit can pass saddle point
SP2, and a part of the norm (power) is transferred to site $1$, as shown in
Fig. \ref{Fig3}(b).

If the incident wave comes from the left, it can be considered as the
perturbation added to the stationary TDB at site $1$. Then, one arrives at
similar conclusions. First, if the perturbation is small, no transport takes
place, as shown in Fig. \ref{Fig3}(c). Next, if the perturbation is large,
leading to $H>E_{\mathrm{thr}1}$, a part of the norm (power) is transferred
to site $3$, as shown in Fig. \ref{Fig3}(d).

Note that in Figs. \ref{Fig3}(b) and \ref{Fig3}(c) we have chosen
identical initial conditions, i.e., there are the same energies of
the TDB, $H$, and the same incident excitations, coming from the
opposite directions. Thus, it follows from the above
considerations that, in the intermediate case,
\begin{equation}
E_{\mathrm{thr}2}<H<E_{\mathrm{thr}1},  \label{<<}
\end{equation}%
the transport from one side to the other can take place in Fig. \ref{Fig3}%
(b), while it is forbidden in Fig. \ref{Fig3}(c). In other words, the same
incident wave can pass from right to left, but \emph{not} in the opposite
direction, which implies non-reciprocal transmission.

\begin{figure}[tbp]
\includegraphics[width=8.0cm]{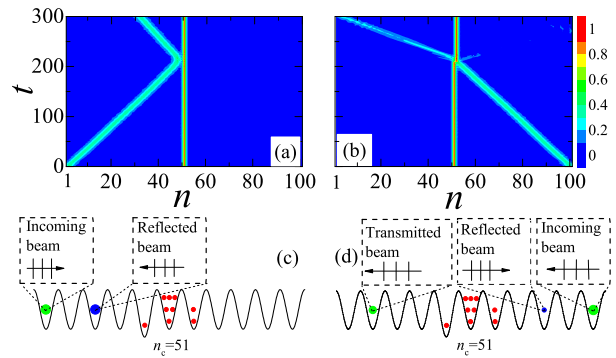}
\caption{The transfer of incident waves through the stationary TDB
in the long lattice ($M=101$ sites), for identical excitations
arriving from the left in (a) and from the right in (b),
respectively. Colors represent norms at the sites. In (c) and (d),
the small and red circles depict the TDBs located,
essentially, at sites $50,51$, and $52$, with $A_{50}=0.1502$, $%
A_{51}=0.974553$, $A_{52}=0.166384$. Circles of different size and color
represent the incident, reflected, and transmitted excitations. The
amplitudes of the incident excitations are $A_{1}=A_{4}=0.2$ and $%
A_{2}=A_{3}=0.4$, and its energy is $\protect\delta E=0.39743$. The energy
thresholds are $E_{\mathrm{thr}1}=-1.14325$ in (a) and $E_{\mathrm{thr}%
2}=-1.30276$ in (b) for the excitations arriving from the left and
right,
respectively. In all the cases, $\protect\lambda =3$, $\protect\varepsilon %
_{n}=0$, except for $\protect\varepsilon _{50}=-0.3$, and $E_{\mathrm{TDB}%
}=-1.6573$.}
\label{Fig4}
\end{figure}

\begin{figure}[tbp]
\includegraphics[width=8.5cm]{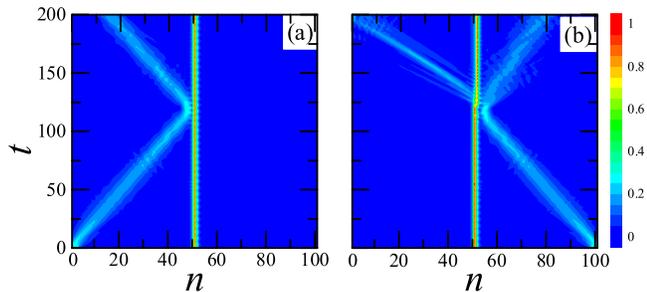}
\caption{The transfer of incident waves through the stationary TDB
in the long lattice ($M=101$ sites) with weaker intrinsic
nonlinearity of identical excitations arriving from the left in
(a) and the right in (b), respectively (cf. Fig.
\protect\ref{Fig4} where the effective nonlinearity of the
incident pulses is stronger by a factor $\approx 3$). The TDB is
located, essentially, at sites $50,51$, and $52$. The amplitudes
of the incident excitations are $A_{1}=A_{4}=0.16$ and
$A_{2}=A_{3}=0.32$. In all the cases,
$\protect\lambda =2.5$, $\protect\varepsilon _{n}=0$, except for $\protect%
\varepsilon _{50}=-0.8$, $\protect\varepsilon _{52}=0.5$.}
\label{SFig3}
\end{figure}

\begin{figure*}[tbp]
\includegraphics[width=16cm]{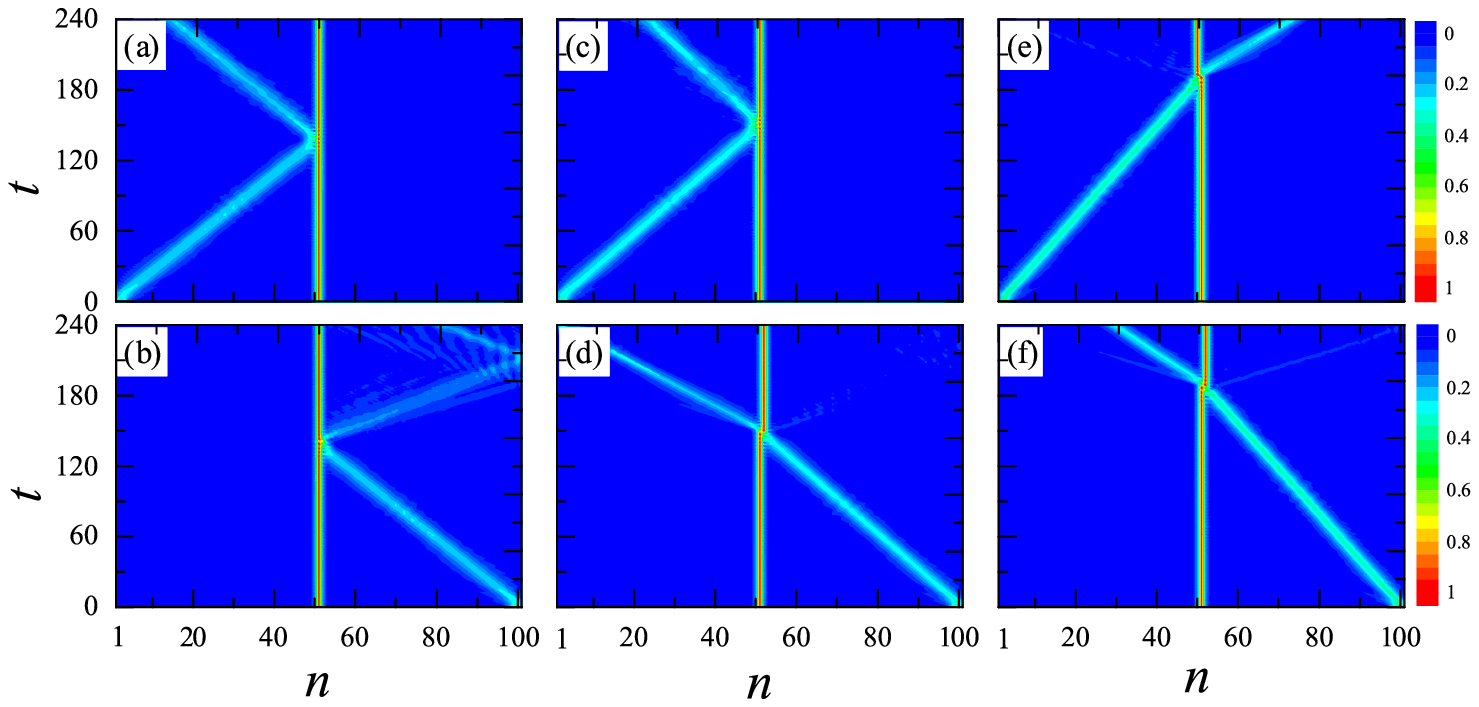}
\caption{The transfer of incident waves through the stationary TDB in the
long lattice ($M=101$ sites) with different energies. Colors represent norms
at the sites. The amplitudes of the incident excitations are $%
A_{2}=A_{3}=0.32$, and its energy is $\protect\delta
E=0.2594<0.271$ in (a) and (b); $A_{2}=A_{3}=0.35$ and
$0.271<\protect\delta E=0.2966<0.315$ in (c) and (d);
$A_{2}=A_{3}=0.4$ and $\protect\delta E=0.38395>0.315$ in (e) and
(f). In all the cases, $A_{1}=A_{4}=0.18$, $\protect\lambda =3$,
$\protect\varepsilon _{n}=0$, except for $\protect\varepsilon
_{50}=-0.05$.} \label{SFigadd}
\end{figure*}

Next, we have performed systematic simulations of the transport of
the incident waves in the full extended lattice of size $M=101$
with the embedded TDB. The first typical example is displayed in
Fig. \ref{Fig4}, where the incident waves arrive from the left or
right to collide with the TDB, as shown in Figs. \ref{Fig4}(c) and
\ref{Fig4}(d), respectively. Initially, the TDB is taken in the
stationary form predicted by the trimer model, the
respective energy thresholds being $E_{\mathrm{thr}1}=-1.14325$ and $E_{%
\mathrm{thr}2}=-1.30276$, with $\lambda =3$, $\varepsilon _{1}=-0.3$, $%
\varepsilon _{3}=0$, and $H_{\mathrm{TDB}}=-1.6573$. The initial condition
for the incident wave packet was set on four sites of the lattice, $\psi _{%
\mathrm{in}}(t)=\{A_{n}e^{i\theta _{n}},~n=1,2,3,4\},$ with phase shifts
corresponding to the staggered ansatz, $\delta \theta _{12}=\delta \theta
_{23}=\delta \theta _{34}=\pi $, where $\delta \theta _{jk}=\theta
_{j}-\theta _{k}$, cf. Eq. (\ref{123}). Its energy $\delta E=\sum_{n=1}^{4}%
\frac{\lambda }{2}A_{n}^{4}+\sum_{n=1}^{3}A_{n}A_{n+1}\cos \theta
_{n,n+1}=0.39743$\textbf{, } hence the total energy, including the TDB and
the incident wave, is
\begin{equation}
E_{t}=H_{\mathrm{TDB}}+\delta E.  \label{delta}
\end{equation}%
The most interesting case for the simulations is
\begin{equation}
E_{\mathrm{thr}2}<E_{t}<E_{\mathrm{thr}1},  \label{EEE}
\end{equation}
as Eq. (\ref{<<}) suggests that the unidirectional transfer may be
expected in this case [parameters of Fig. \ref{Fig4} comply with
Eq. (\ref{EEE})]. The simulations confirm this expectation: while
the wave arriving from the left is entirely reflected by the TDB,
as seen in Figs. \ref{Fig4}(a) and \ref{Fig4}(c), the one coming
from the right is chiefly transmitted through the TDB, although a
weak component is reflected, as seen in Figs. \ref{Fig4}(b) and
\ref{Fig4}(d).

Note that, as seen in Fig. \ref{Fig4}(b), the impact of the
incident packet which passes the TDB leads to a shift of the TDB
by one site in the direction from which the packet has arrived.
The shift actually
takes place \emph{against} the lattice-potential slope, as seen from Eq. (%
\ref{asymm}). The shift, which is explained by the mutual nonlinear
attraction between the incident and pinned modes, is, as a matter of fact, a
manifestation of the \textit{tractor-beam} effect, that has recently drawn
much interest in optics \cite{tractor}. On the other hand, Fig. \ref{Fig4}%
(a) shows that the reflected wave packet does not shift the pinned TDB,
which is explained by the balance of the attraction and recoil effects.

If the lattice contains only isolated defects embedded into a
regular host medium, the shifts produced by single or multiple
scatterings of wave packets may draw the TDB into a uniform
section, where the diode effect will disappear. On the other hand,
if the lattice is built as a chain of defects of type
(\ref{asymm}), i.e., weak asymmetry is present at each site, the
diode and tractor-beam effects may possibly persist in multiple
collisions of incident wave packets with the TDB, until the
gradually drifting TDB will hit the edge of the system (the left
edge, if the wave packets arrive from the left). Of course, the
mean potential slope in such a tilted lattice will eventually give
rise to reflection of any incident wave packet, in the absence of
the TDB; however, the respective reflection length is much larger
than that induced by the TDB, hence the diode effect may plausibly
remain discernible in the tilted lattice.

In Fig. \ref{Fig4}, the examples of the transmission and
reflection are demonstrated for incident wave packets which carry
relatively strong nonlinearity. Similar results for incident wave
packets with the strength of the intrinsic nonlinearity weaker by
a factor of $3$ than in Fig. \ref{Fig4} are displayed in Fig.
\ref{SFig3}. It is seen that the diode effect acts in this case
too. On the other hand, the effect cannot take place for linear
wave packets (ones with a very small amplitude), as, having small
energies, they do not satisfy condition (\ref{EEE}).

Another set of characteristic examples is plotted in Fig. \ref{SFigadd},
where, keeping fixed parameters of the defect and TDB pinned to it, we
illustrate the change of the scattering scenario with the increase of energy
$\delta E$ of the incident wave packet. As predicted by the above analysis,
at small energies, corresponding to $E_{t}<E_{\mathrm{thr}2}$, see Eq. (\ref%
{delta}), Figs. \ref{SFigadd}(a) and \ref{SFigadd}(b) corroborate
that the wave packets cannot pass the TDB in either direction. On
the other hand, the diode and tractor-beam effects act in Figs.
\ref{SFigadd}(c) and \ref{SFigadd}(d), which was generated for
$\delta E$ complying with Eq. (\ref{EEE}). Lastly, Figs.
\ref{SFigadd}(e) and \ref{SFigadd}(f), which corresponds to
$E_{t}>E_{\mathrm{thr}1}$, demonstrates that left- and
right-arriving wave packets easily pass the TDB. In the latter
case, the ``tractor-beam" effect takes place for either incidence
direction.

Collecting results of the simulations for the full lattice, we conclude that
$\delta E$ may be used as an efficient control parameter to govern the
outcome of the collision of the incident wave packet with the given TDB, in
accordance with Eq. (\ref{EEE}). To clearly display the results, in Fig. \ref%
{Fig6} we present the reflection, transmission, and absorption coefficients,
defined in terms of the energy, for the passage of the left- and
right-incident wave packets through the given TDB versus $\delta E$ in the
long lattice ($M=101$ sites), ``absorption" meaning that a
part of the energy of the incident wave may be spent on weak excitation of
the quiescent TDB. Here, the initial conditions similar to those in Fig. \ref%
{SFigadd}, with the incident excitation having $A_{1}=A_{4}=0.18$,
while its energy, $\delta E$, is adjusted by varying the value of $A_{2}=A_{3}$.%
The sum of the three coefficients is $1$, as it must be. In region I, $%
E_{t}<E_{\mathrm{thr}2}<E_{\mathrm{thr}1}$, the transmission
coefficients for both the left- and right-incident waves are
practically zero. In the middle region II, which corresponds to
Eq. (\ref{EEE}), only the right-incident wave passes the TDB,
while the left wave is reflected, i.e., the TDB-induced diode
effect takes place. The effect can be controlled by adjusting
$\varepsilon _{n-1}$ and $\varepsilon _{n+1}$. Finally, in region
III, $E_{\mathrm{thr}2}<E_{\mathrm{thr}1}<E_{t}$, the TDB is
passable in both directions. In particular, if $\varepsilon
_{n-1}\rightarrow -\infty $, SP1 disappears [see Fig.
\ref{SFig1}(a) in the Appendix], and region III does not exist. In
the latter case, the incident wave packets cannot pass the TDB
from left to right, irrespective of $\delta E$, and the TDB tends
to acts as a ``full diode".

\begin{figure}[tbp]
\includegraphics[width=8.0cm]{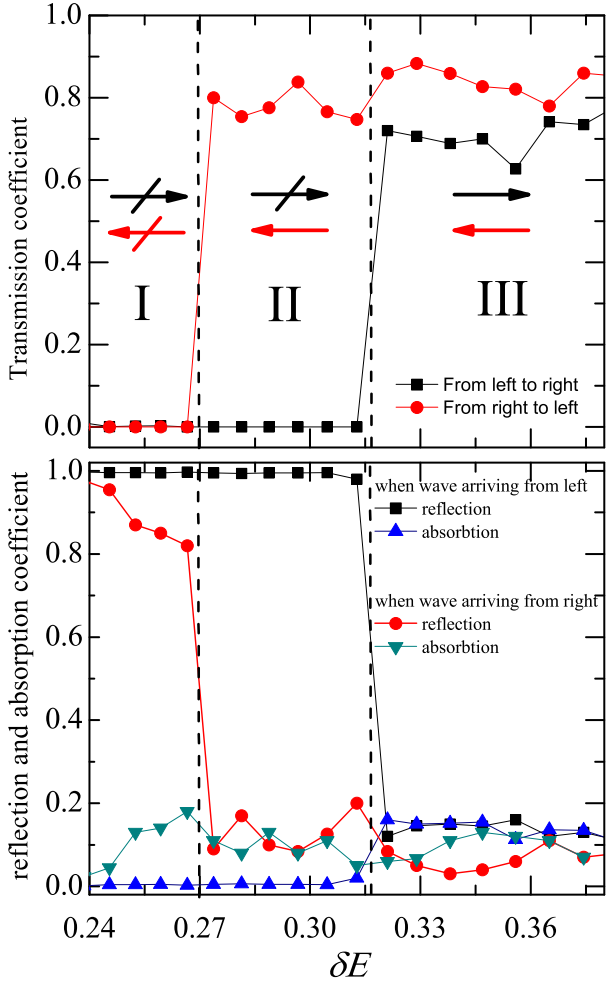}
\caption{ The transmission, reflection, and absorption
coefficients for the passage through the stationary TDB in the
long lattice ($M=101$ sites) of the same incident wave packet
arriving from the left and from the right, as functions of its
energy, $\protect\delta E$. The energy thresholds for the
excitation coming from the left and right satisfy $E_{\mathrm{thr}1}>E_{\mathrm{%
thr}2}$, and $\rightarrow $ ($\nrightarrow $) implies that the incident wave
can (cannot) pass the TDB. Regions I, II, and III are defined by,
respectively, $E_{t}<E_{\mathrm{thr}2}$, $E_{\mathrm{thr}2}<E_{t}<E_{\mathrm{%
thr}1}$, and $E_{t}>E_{\mathrm{thr}1}$. The TDB is located, essentially, at
sites $50,51$, and $52$, with $A_{50}=0.163359$, $A_{51}=0.972447$, and $%
A_{52}=0.166316$. In all the cases, $\protect\lambda =3$ and $\protect%
\varepsilon _{n}=0$, except for $\protect\varepsilon _{50}=-0.05$.}
\label{Fig6}
\end{figure}

\section{Conclusion}

\label{sec4} We have investigated the transport of waves through
the TDB in the framework of the DNLS model with the asymmetric
on-site defect potential. As a result, the diodelike transport
mode has been found, i.e., the unidirectional transfer of the
waves across the TDB, in a finite interval of energies of the
incident wave packet, while the underlying lattice itself (without
the TDB placed at a desired position) cannot operate in such a
regime. The underlying mechanism is accurately explained by the
consideration of respective energy barriers. If the incident wave
packet passes the pinned TDB, the ``tractor-beam" effect takes
place, with the TDB shifting by one lattice site in the incidence
direction, due to its attraction to the incident pulse. The
bouncing wave packet does not cause the shift, the attraction
being balanced by the recoil.

In the experimental realization of the system in BEC, the defect potential
can be created, respectively, as a barrier or well, by a blue- or
red-detuned laser beam illuminating the BEC at a particular site \cite%
{ATrombettoni2002,JELye2005,CFort2005}. In terms of optical waveguide
arrays, the potential can be induced by altering the effective refractive
index in particular cores. The results suggest a scheme for the
implementation of controlled blocking, filtering, and routing of matter-wave
and optical beams in guiding networks, as well as for the realization of the
tractor-beam mechanism. In particular, the scheme may be useful for steering
matter waves in interferometry and quantum-information processing \cite{RAVicencio2007}. In addition to working with the atom and optical beams,
the present results may find applications in various other contexts to which
the DNLS model applies.

\section*{Acknowledgments}

\label{sec5} This work is supported by the National Natural
Science Foundation of China under Grants No. 11474026 and No. 11674033, and the
Fundamental Research Funds for the Central Universities under
Grant No. 2015KJJCA01. The work of B.A.M. is supported, in part,
by Grant No. 2015616 from the joint program in physics between NSF
and Binational (US-Israel) Science Foundation.

\section*{Appendix: The consideration of limit cases}
\label{sec6}

In the main text, we mainly consider the passage of wave
packets in the opposite directions through the TDB \ pinned by the
asymmetric local potential. As a result, we have found that the
unidirectional passage (the diode effect) strongly depends on the total
energy, $E_{t}$, and the threshold energies, $E_{\mathrm{thr}1,2}$. As shown
in Fig. \ref{Fig6} in the main text, the TDB is impassable for the wave
packets arriving from either side in the case of $E_{t}<E_{\mathrm{thr}2}<E_{%
\mathrm{thr}1}$, which is defined as region I, and passable in region III, $%
E_{\mathrm{thr}2}<E_{\mathrm{thr}1}<E_{t}$. On the other hand, in region II,
which corresponds to $E_{\mathrm{thr}2}<E_{t}<E_{\mathrm{thr}1}$, the
incident wave passes the TDB unidirectionally (from right to left in Fig. %
\ref{Fig4} of the main text), with a high transmission
coefficient. Here, we discuss limit cases in which the absolute
value of $\protect\varepsilon _{1}$ is large enough.

\begin{figure}[!ht]
\begin{center}
\includegraphics[width=8.0cm]{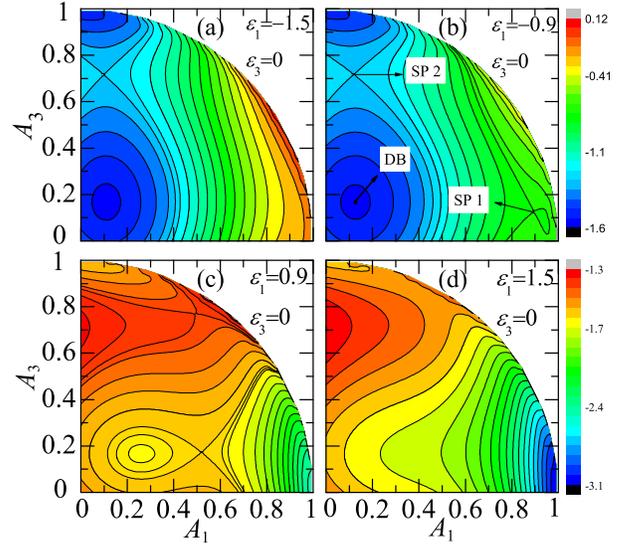}
\caption{Contour plots of the energy given by Eq. (8) from the main text
when the absolute value of $\protect\varepsilon _{1}$ is large enough, with
fixed $\protect\lambda =3$ and $\protect\varepsilon _{3}=0$. SP1 and SP2
denote the saddle points.}
\label{SFig1}
\end{center}
\end{figure}

In the main text it has been demonstrated that the left- (right-) incident
wave packet passes the TDB only when its energy exceeds $E_{\mathrm{thr}1}$ ($E_{\mathrm{thr}2}$). That is, the TDB creates different potential barriers $E_{\mathrm{thr}1}$ and $E_{\mathrm{thr}2}$ for the same wave packet arriving
from the different directions. The effect of coefficients $\varepsilon _{n}$, that determine the local defect, on the Hamiltonian, considered
as a function of $A_{1}$ and $A_{3}$, is shown in Figs.
\ref{Fig2}(a) and \ref{Fig2}(c) in the main text. When
$\varepsilon _{3}=0$ and $\varepsilon _{1}=-0.7$, the discrete
soliton is a TDB, with $E_{\mathrm{thr}1}>E_{\mathrm{thr}2}$. In
this case, although it is harder for the wave packet to pass the
TDB from left to right, it still can do that at
$E_{t}>E_{\mathrm{thr}1}$. However, when $\varepsilon _{1}$ is
negative, and its absolute is large, the energy at saddle point
SP1 is much higher than at SP2, i.e., $E_{\mathrm{thr}1}\gg
E_{\mathrm{thr}2}$, as shown in Fig. \ref{SFig1}(b). In
particular, when $|\varepsilon _{1}|$ is large enough, SP1
disappears, and value $E_{\mathrm{thr}1}$ cannot be reached, as
shown in Fig. \ref{SFig1}(a) (cf. the consideration of the trimer
system in\textbf{\ }Ref. \cite{GKopidakis2000}). Effectively, in
the latter case, the TDB is an infinitely high potential barrier
for the wave packet arriving from the left. Hence, irrespective of
the magnitude of $E_{t}$, the wave packet cannot pass the TDB from
left to right. (as long as the norm of the scattering wave is not
too large compared to the norm of the scattering TDB). On the
other hand, when $\varepsilon _{1}$ is positive and large enough,
SP1 and DB disappear. Accordingly, thresholds
$E_{\mathrm{thr}1,2}$ lose their meaning, as shown in Figs.
\ref{SFig1}(c) and \ref{SFig1}(d). Note that the transfer of the
wave packet through the TDB from right to left can be controlled
by means of potential parameter $\varepsilon _{3}$.

\begin{thebibliography}{99}

\bibitem{AJSievers1998} A. J. Sievers and J. B. Page, in Phonon Physics: The
Cutting Edge, Dynamical Properties of Solids Vol. VII (Elsevier,
Amsterdam, 1995); S. Aubry, Physica (Amsterdam) \textbf{103}D, 201
(1997); S. Flach and A. V. Gorbach, Phys. Rep. \textbf{467}, 1
(2008); J. W. Fleischer, M. Segev, N. K. Efremidis, and D. N.
Christodoulides, Nature (London) \textbf{422}, 147 (2003).


\bibitem{SFlach1998} S. Flach and C. R. Willis, Phys. Rep. \textbf{295}, 181
(1998).


\bibitem{Panos} P. G. Kevrekidis, The Discrete Nonlinear Schr\"{o}dinger
Equation: Mathematical Analysis, Numerical Computations, and Physical
Perspectives (Berlin: Springer, 2009).


\bibitem{MSato2006} M. Sato, B. E. Hubbard, and A. J. Sievers, Rev. Mod.
Phys. \textbf{78}, 137 (2006).


\bibitem{ETrias2000} E. Trias, J. J. Mazo, and T. P. Orlando, Phys. Rev.
Lett. \textbf{84}, 741 (2000).


\bibitem{AVUstinov2003} A. V. Ustinov, Chaos \textbf{13}, 716 (2003).


\bibitem{DNChr} D. N. Christodoulides and R. I. Joseph, Opt. Lett. \textbf{13}, 794 (1988).

\bibitem{RMorandotti1999} R. Morandotti, U. Peschel, J. S. Aitchison, H. S.
Eisenberg, and Y. Silberberg, Phys. Rev. Lett. \textbf{83}, 2726 (1999).


\bibitem{HSEisenberg1998} F. Lederer, G. I. Stegeman, D. N. Christodoulides,
G. Assanto, M. Segev, and Y. Silberberg, Phys. Rep. \textbf{463}, 1 (2008);
Y. V. Kartashov, V. A. Vysloukh, and L. Torner, Progr. Opt. \textbf{52}, 63
(2009); U. R\"{o}pke, H. Bartelt, S. Unger, K. Schuster, and J. Kobelke,
Appl. Phys. B\textbf{\ 104}, 481 (2011).


\bibitem{Kobelke} F. Eilenberger, S. Minardi, A Szameit, U. R\"{o}pke, J.
Kobelke, K. Schuster, H. Bartelt, S. Nolte, A. T\"{u}nnermann, and T.
Pertsch, Opt. Express \textbf{19}, 23171 (2011).


\bibitem{UTSchwarz1999} U. T. Schwarz, L. Q. English, and A. J. Sievers,
Phys. Rev. Lett. \textbf{83}, 223 (1999).


\bibitem{MSato2004} M. Sato and A. J. Sievers, Nature (London) \textbf{432},
486 (2004).


\bibitem{ATrombettoni2001} A. Trombettoni and A. Smerzi, Phys. Rev. Lett.
\textbf{86}, 2353 (2001); F. K. Abdullaev, B. B. Baizakov, S. A. Darmanyan,
V. V. Konotop, and M. Salerno, Phys. Rev. A \textbf{64}, 043606 (2001); R.
Carretero-Gonz\'{a}lez and K. Promislow, Phys. Rev. A \textbf{66}, 033610
(2002); G. I. Alfimov, P. G. Kevrekidis, V. V. Konotop, and M. Salerno,
Phys. Rev. E \textbf{66}, 046608 (2002).


\bibitem{BEiermann2004} B. Eiermann, T. Anker, M. Albiez, M. Taglieber, P.
Treutlein, K. P. Marzlin, and M. K. Oberthaler, Phys. Rev. Lett. \textbf{92}, 230401 (2004); O. Morsch and M. Oberthaler, Rev. Mod. Phys. \textbf{78},
179 (2006).


\bibitem{Konotop} V. A. Brazhnyi and V. V. Konotop, Mod. Phys. Lett. B
\textbf{18}, 627 (2004).


\bibitem{Kolomeisky1992} E. B. Kolomeisky and J. P. Straley, Phys. Rev. B
\textbf{46}, 11749 (1992); E. B. Kolomeisky, T. J. Newman, J. P. Straley,
and X. Qi, Phys. Rev. Lett. \textbf{85}, 1146 (2000).


\bibitem{JuKuiXue2008} J. K. Xue and A. X. Zhang, Phys. Rev. Lett. \textbf{%
101}, 180401(2008); A. X. Zhang and J. K. Xue, Phys. Rev. A \textbf{80},
043617 (2009).


\bibitem{MGClerc2011} M. G. Clerc, R. G. Elias, and R. G. Rojas, Phil.
Trans. R. Soc. London Ser. B \textbf{369}, 412 (2011).


\bibitem{PCMatthews2011} P. C. Matthews and H. Susanto, Phys. Rev. E \textbf{%
84}, 066207 (2011).


\bibitem{JDAndersen1993} J. D. Andersen and V. M. Kenkre, Phys. Rev. B
\textbf{47}, 11134 (1993).


\bibitem{MPeyrard1993} M. Peyrard, T. Dauxois, H. Hoyet, and C. R. Willis,
Physica D \textbf{68}, 104 (1993).


\bibitem{RSMacKay1998} R. S. MacKay and J. A. Sepulchre, Physica D \textbf{%
119}, 148 (1998).


\bibitem{PJMartinez2003} P. J. Martinez, M. Meister, L. M. Floria, and F.
Falo, Chaos \textbf{13}, 610 (2003).


\bibitem{GPTsironis1996} G. P. Tsironis and S. Aubry, Phys. Rev. Lett.
\textbf{77}, 5225 (1996).


\bibitem{RLivi2006} R. Livi, R. Franzosi, and G. L. Oppo, Phys. Rev. Lett.
\textbf{97}, 060401 (2006).


\bibitem{GSNg2009} G. S. Ng, H. Hennig, R. Fleischmann, T. Kottos, and T.
Geisel, New J. Phys. \textbf{11}, 073045 (2009).


\bibitem{RFranzosi2007} R. Franzosi, R. Livi, and G.-L. Oppo, J. Phys. B
\textbf{40}, 1195 (2007).


\bibitem{XDBai2012} X.-D. Bai and J.-K. Xue, Phys. Rev. E \textbf{86},
066605 (2012). X.-D. Bai, A.-X. Zhang, and J.-K. Xue \textbf{88}, 062916
(2013).


\bibitem{TSchumm2005} T. Schumm, S. Hofferberth, L. M. Andersson, S.
Wildermuth, S. Groth, I. Bar-Joseph, J. Schmiedmayer, and P. Kr\"{u}ger,
Nat. Phys. \textbf{1}, 57 (2005).


\bibitem{DJaksch1999} D. Jaksch, H. J. Briegel, J. I. Cirac, C. W. Gardiner,
and P. Zoller, Phys. Rev. Lett. \textbf{82}, 1975 (1999).


\bibitem{GKBrennen1999} G. K. Brennen, C. M. Caves, P. S. Jessen, and I. H.
Deutsch, Phys. Rev. Lett.\ \textbf{82}, 1060 (1999).


\bibitem{JKPachos2003} J. K. Pachos and P. L. Knight, Phys. Rev. Lett.
\textbf{91}, 107902 (2003).


\bibitem{A.Kay2006} A. Kay, J. K. Pachos, and C. S. Adams, Phys. Rev. A
\textbf{73}, 022310 (2006).


\bibitem{Lepri} S. Lepri and G. Casati, Phys. Rev. Lett. \textbf{106},
164101 (2011); S. Lepri and B. A. Malomed, Phys. Rev. E \textbf{87}, 042903
(2013).


\bibitem{Assanto} K. Gallo, G. Assanto, K. R. Parameswaran, and M. M. Fejer,
Appl. Phys. Lett. \textbf{79}, 314 (2001).


\bibitem{Bragg-uni} M. Kulishov, J. M. Laniel, N. B\'{e}langer, J. Aza\~{n}%
a, and D. V. Plant, Opt. Express \textbf{13}, 3068 (2005); S. Ding
and G. P. Wang, Appl. Phys. Lett. \textbf{100}, 151913 (2012).


\bibitem{meta} M. W. Feise, I. V. Shadrivov, and Y. S. Kivshar, Phys. Rev. E
\textbf{71}, 037602 (2005).


\bibitem{atomic-chain-uni} J.-H. Wu, M. Artoni, and G. C. La Rocca, Phys.
Rev. Lett. \textbf{113}, 123004 (2014); C. Sayrin, C. Junge, R. Mitsch, B.
Albrecht, D. O'Shea, P. Schneeweiss, J. Volz, and A. Rauschenbeutel, Phys.
Rev. X \textbf{5}, 041036 (2015).


\bibitem{Fabio} V. Grigoriev and F. Biancalana, Opt. Lett. \textbf{36}, 2131
(2011).


\bibitem{Scalora} M. Scalora, J. P. Dowling, C. M. Bowden, and M. J.
Bloemer, J. Appl. Phys. \textbf{76}, 2023 (1994).


\bibitem{Ablowitz} M. J. Ablowitz, C. W. Curtis, and Y.-P. Ma, Phys. Rev. A
\textbf{90}, 023813 (2014); Z. Li, R.-X. Wu, Y. Poo, Z. F. Lin, and Q. B.
Li, J. Optics \textbf{16}, 125004 (2014).


\bibitem{PhotCryst-uni} S. Feng and Y. Wang, Optical Materials \textbf{35},
1455 (2013); Opt. Exp. \textbf{21}, 220 (2013); A. Cicek and B. Ulug, Appl.
Phys. B - Lasers and Opt. \textbf{113}, 619 (2013); L. H. Wang, X. L. Yang,
X. F. Meng, Y. R. Wang, S. X. Chen, Z. Huang, and G. Y. Dong, Jpn. J. Appl.
Phys. \textbf{52}, 122601 (2013); P. Wang, C. Ren, P. Han, and S. Feng, Opt.
Mater. \textbf{46}, 195 (2015).


\bibitem{VKVK} V. V. Konotop and V. Kuzmiak, Phys. Rev. B \textbf{66}, 14
(2002).


\bibitem{Fabio2} F. Biancalana, J. Appl. Phys. \textbf{104}, 093113 (2008).


\bibitem{PT-uni} H. Ramezani, T. Kottos, R. El-Ganainy, and D. N.
Christodoulides, Phys. Rev. A \textbf{82}, 043803 (2010); A.
Regensburger, C. Bersch, M. A. Miri, G. Onishchukov, D. N.
Christodoulides, and U. Peschel, Nature \textbf{488}, 167 (2012);
M.-A. Miri, A. Regensburger, U. Peschel, and D. N.
Christodoulides, Phys. Rev. A \textbf{86}, 023807 (2012); J.
D'Ambroise, P. G. Kevrekidis, and S. Lepri, J. Phys. A: Math.
Theor. \textbf{45}, 444012 (2012); J. D'Ambroise, S. Lepri, B. A.
Malomed, and P. G. Kevrekidis, Phys. Lett. A \textbf{378}, 2824
(2014); J. Gear, F. Liu, S. T. Chu, S. Rotter, and J. Li, Phys.
Rev. A. \textbf{91}, 033825 (2015); S. Yu, X. Piao, K. W. Yoo, J.
Shin, and N. Park, Opt. Express \textbf{23}, 24997 (2015); Y. Jia,
Y. Yan, S. V. Kesava, E. D. Gomez, and N. C. Giebink, ACS
Photonics \textbf{2}, 319 (2015);


\bibitem{micro-uni} S. Kirihara, M. W. Takeda, K. Sakoda, and Y. Miyamoto,
Solid State Comm. \textbf{124}, 135 (2002).


\bibitem{electric-uni} F. Tao, W. Chen, W. Xu, J. T. Pan, and S. D. Du,
Phys. Rev. E \textbf{83}, 056605 (2011).


\bibitem{PWAnderson1958} P. W. Anderson, Phys. Rev. \textbf{109}, 1492
(1958).


\bibitem{JBilly2008} J. Billy, V. Josse, Z. Zuo, A. Bernard, B. Hambrecht,
P. Lugan, D. Clement, L. Sanchez-Palencia, P. Bouyer, and A. Aspect, Nature
(London) \textbf{453}, 891 (2008).


\bibitem{GKopidakis2000} G. Kopidakis and S. Aubry, Physica D \textbf{130},
155 (1999); G. Kopidakis and S. Aubry, Physica D \textbf{139}, 247 (2000);
G. Kopidakis and S. Aubry, Phys. Rev. Lett. \textbf{84}, 3236 (2000).


\bibitem{LFallani2007} L. Fallani, J. E. Lye, V. Guarrera, C. Fort, and M.
Inguscio, Phys. Rev. Lett. \textbf{98}, 130404 (2007).


\bibitem{AIomin2010} M. I. Molina, Phys. Rev. B \textbf{58}, 12547 (1998);
A. S. Pikovsky and D. L. Shepelyansky, Phys. Rev. Lett. \textbf{100}, 094101
(2008); S. Flach, D. O. Krimer, and Ch. Skokos, \textit{ibid}.. \textbf{102}%
, 024101 (2009); Ch. Skokos, D. O. Krimer, S. Komineas, and S. Flach, Phys.
Rev. E \textbf{79}, 056211 (2009); H. Veksler, Y. Krivolapov, and S.
Fishman, \textit{ibid}. \textbf{80}, 037201 (2009); A. Iomin, \textit{ibid}.
E \textbf{81}, 017601 (2010); E. Lucioni, B. Deissler, L. Tanzi, G. Roati,
M. Zaccanti, M. Modugno, M. Larcher, F. Dalfovo, M. Inguscio, and G.
Modugno, Phys. Rev. Lett. \textbf{106}, 230403 (2011); B. Min, T. Li, M.
Rosenkranz, and W. Bao, Phys. Rev. A \textbf{86}, 053612 (2012).


\bibitem{MJohansson2010} M. Johansson, G. Kopidakis, and S. Aubry, Europhys.
Lett. \textbf{91}, 50001 (2010).


\bibitem{ZYSun2015} Z.-Y. Sun and S. Fishman, Phys. Rev. E \textbf{92},
040903(R) (2015).


\bibitem{DJaksch1998} D. Jaksch, C. Bruder, J. I. Cirac, C. W. Gardiner, and
P. Zoller, Phys. Rev. Lett. \textbf{81}, 3108 (1998).


\bibitem{HHennig2010} H. Hennig, J. Dorignac, and D. K. Campbell, Phys. Rev.
A \textbf{82}, 053604 (2010).


\bibitem{tractor} A. Novitsky, C.-W. Qiu, and H. Wang, Phys. Rev. Lett.
\textbf{107}, 203601 (2011); S. Sukhov and A. Dogariu, \textit{ibid}.
\textbf{107}, 203602 (2011); O. Brzobohaty, V. Karasek, M. Siler, L.
Chvatal, T. Cizmar, and P. Zemanek, Nature Phot. \textbf{7}, 123 (2013); V.
Shvedov, A. R. Davoyan, C. Hnatovsky, N. Engheta, and W. Krolikowski,
\textit{ibid}. \textbf{8}, 846 (2014); H. Chen, S. Liu, J. Zi, and Z. F.
Lin, ACS Nano \textbf{9}, 1926 (2015).


\bibitem{CFort2005} C. Fort, L. Fallani, V. Guarrera, J. E. Lye, M. Modugno,
D. S. Wiersma, and M. Inguscio, Phys. Rev. Lett. \textbf{95}, 170410 (2005).


\bibitem{JELye2005} J. E. Lye, L. Fallani, M. Modugno, D. S. Wiersma, C.
Fort, and M. Inguscio, Phys. Rev. Lett. \textbf{95}, 070401 (2005).


\bibitem{ATrombettoni2002} A. Trombettoni, A. Smerzi, and A. R. Bishop,
Phys. Rev. Lett. \textbf{88}, 173902 (2002)


\bibitem{RAVicencio2007} R. A. Vicencio, J. Brand, and S. Flach, Phys. Rev.
Lett. \textbf{98}, 184102 (2007).


\end{thebibliography}
\end{document}